\documentclass[reprint,amsmath,amssymb, aps,pre,floatfix,superscriptaddress,longbibliography]{revtex4-1}

\usepackage{graphicx} 
\usepackage[separate-uncertainty=true]{siunitx}

\DeclareSIUnit\molar{M}
\DeclareSIUnit{\wtpc}{wt\%}
\usepackage[colorlinks,allcolors=blue]{hyperref}

\usepackage{comment}
\usepackage{xcolor}
\usepackage{bm}

\usepackage{tikz}
\usepackage{scalerel}
\usetikzlibrary{svg.path}

\definecolor{deep0}{rgb}{0.30,0.45,0.69}
\definecolor{deep1}{rgb}{0.87,0.52,0.32}
\definecolor{deep2}{rgb}{0.33,0.66,0.41}
\definecolor{deep3}{rgb}{0.77,0.31,0.32}
\definecolor{deep4}{rgb}{0.51,0.45,0.70}
\newcommand{\plb}[1]{\textbf{(\MakeLowercase{#1})}} 

\makeatletter
\def\maketitle{
\@author@finish
\title@column\titleblock@produce
\suppressfloats[t]}
\makeatother

\AtBeginDocument{%
    \newwrite\bibnotes
    \def\bibnotesext{Notes.bib}
    \immediate\openout\bibnotes=\jobname\bibnotesext
    \immediate\write\bibnotes{@CONTROL{REVTEX41Control}}
    \immediate\write\bibnotes{@CONTROL{%
    apsrev41Control,author="08",editor="1",pages="1",title="0",year="1"}}
     \if@filesw
     \immediate\write\@auxout{\string\citation{apsrev41Control}}%
    \fi
}%

\let\vec\mathbf

\newcommand\Pen{\mbox{\textit{Pe}}}  

\definecolor{orcidlogocol}{HTML}{A6CE39}
\tikzset{
  orcidlogo/.pic={
    \fill[orcidlogocol] svg{M256,128c0,70.7-57.3,128-128,128C57.3,256,0,198.7,0,128C0,57.3,57.3,0,128,0C198.7,0,256,57.3,256,128z};
    \fill[white] svg{M86.3,186.2H70.9V79.1h15.4v48.4V186.2z}
                 svg{M108.9,79.1h41.6c39.6,0,57,28.3,57,53.6c0,27.5-21.5,53.6-56.8,53.6h-41.8V79.1z M124.3,172.4h24.5c34.9,0,42.9-26.5,42.9-39.7c0-21.5-13.7-39.7-43.7-39.7h-23.7V172.4z}
                 svg{M88.7,56.8c0,5.5-4.5,10.1-10.1,10.1c-5.6,0-10.1-4.6-10.1-10.1c0-5.6,4.5-10.1,10.1-10.1C84.2,46.7,88.7,51.3,88.7,56.8z};
  }
}

\newcommand\orcid[1]{\href{https://orcid.org/#1}{\mbox{\scalerel*{
\begin{tikzpicture}[yscale=-1,transform shape]
\pic{orcidlogo};
\end{tikzpicture}
}{|}}}}

\makeatletter
\def\convertto#1#2{\strip@pt\dimexpr #2*65536/\number\dimexpr 1#1}
\makeatother

\begin{document}

\newcommand{\goeaffil}{Max Planck Institute for Dynamics and Self-Organization, Am Fa\ss{}berg 17, 37077 G\"ottingen and Institute for the Dynamics of Complex Systems, Georg August Universit\"at G\"ottingen, Germany}
\newcommand{\twaffil}{Physics of Fluids Group, Max Planck Center for Complex Fluid Dynamics, and J. M. Burgers Center for Fluid Dynamics, University of Twente, PO Box 217,7500AE Enschede, Netherlands}
\newcommand{\susaffil}{Southern University of Science and Technology, Shenzhen, 518055, Guangdong, China}
\newcommand{\oldaffil}{University of Oldenburg, Institute of Physics \& ForWind, Küpkersweg 70, 26129 Oldenburg, Germany}

\newcommand{\mytitle}{Magnetotaxis in droplet microswimmers}
\newcommand{\SItitle}{Supporting information: Magnetotaxis in droplet microswimmers}
\title{\mytitle} 
\author{Martin W. Wagner}
\affiliation{\goeaffil}
\affiliation{\oldaffil}

 \author{Freek Domburg}
 \affiliation{\twaffil}

\author{Carsten Kr\"uger}%
\affiliation{\goeaffil}

\author{Jens Meyer}%
\affiliation{\goeaffil}

\author{Jiaqi Zhang}
\affiliation{\susaffil}
\affiliation{\twaffil}

 \author{Prashanth Ramesh~\orcid{0000-0003-3264-5084 }}
\affiliation{\goeaffil}
\affiliation{\twaffil}
 \author{Corinna C. Maass~\orcid{0000-0001-6287-4107}}%
 \email{c.c.maass@utwente.nl}%
\affiliation{\goeaffil}
\affiliation{\twaffil}
\begin{abstract}
Magnetotaxis is a well known phenomenon in swimming microorganisms which sense magnetic fields e.g. by incorporating crystalline magnetosomes. In designing artificial active matter with tunable dynamics, external magnetic fields can provide a versatile method for guidance. Here, it is the question what material properties are necessary to elicit a significant response. In this working paper, we document in  experiments on self-propelling nematic microdroplets that the weak diamagnetic torques exerted by a sub-Tesla magnetic field are already sufficient to significantly affect the dynamics of these swimmers. We find a rich and nontrivial variety of dynamic modes by varying droplet size, state of confinement and magnetic field strength.
\end{abstract}
\date{\today}%
\maketitle
\section*{Introduction}

Microswimmers, i.e.\ autonomously self-propelling agents on micrometric length scales~\cite{marchetti2013_hydrodynamics,bechinger2016_active,moran2019_microswimmers} are an important subclass of active matter, including both biological swimmers like algae and bacteria, as well as their synthetic analogues.

Nature has evolved various strategies to use the Earth's magnetic field for guidance and orientation~\cite{lohmann2007_magnetic}, ranging from magnetoreceptive cues in migratory animals~\cite{wiltschko1995_magnetic,johnsen2005_physics} to simple magnetotaxis in flagellated prokaryotes that produce and incorporate magnetite nanoparticles or magnetosomes~\cite{blakemore1982_magnetotactic}. Particularly the strategies of the latter can be used to inspire sensing, guidance and actuation strategies~\cite{yazdi2018_magnetotaxis,codutti2019_chemotaxis,klumpp2019_swimming,thery2020_selforganisation,thery2024_controlling} in artificial and hybrid active matter and microswimmers~\cite{dreyfus2005_microscopic,medina-sanchez2018_swimming,stanton2017_magnetotactic}. Usually, these strategies require ferrofluids, nanoengineered magnetosomes or magnetic colloids embedded in the swimmer~\cite{fan2020_reconfigurable,dreyfus2005_microscopic,wang2022_order,khan2024_perturbing}.
One well-studied experimental microswimmer system are self-propelling droplets, which have been realized for a number of chemical constituents, including oil-in-water active emulsions where the oil phase can be isotropic or liquid crystalline~\cite{zhang2021_autonomous,wang2021_active,michelin2023_selfpropulsion,babu2022_motile,birrer2022_we}.
In this working paper we report that, in liquid crystalline droplet swimmers, magnetotaxis can be induced by purely diamagnetic interactions.

\section*{Results}

\subsection*{Self propelling nematic droplet swimmers}

\begin{figure}
    \centering
    \includegraphics[width=\columnwidth]{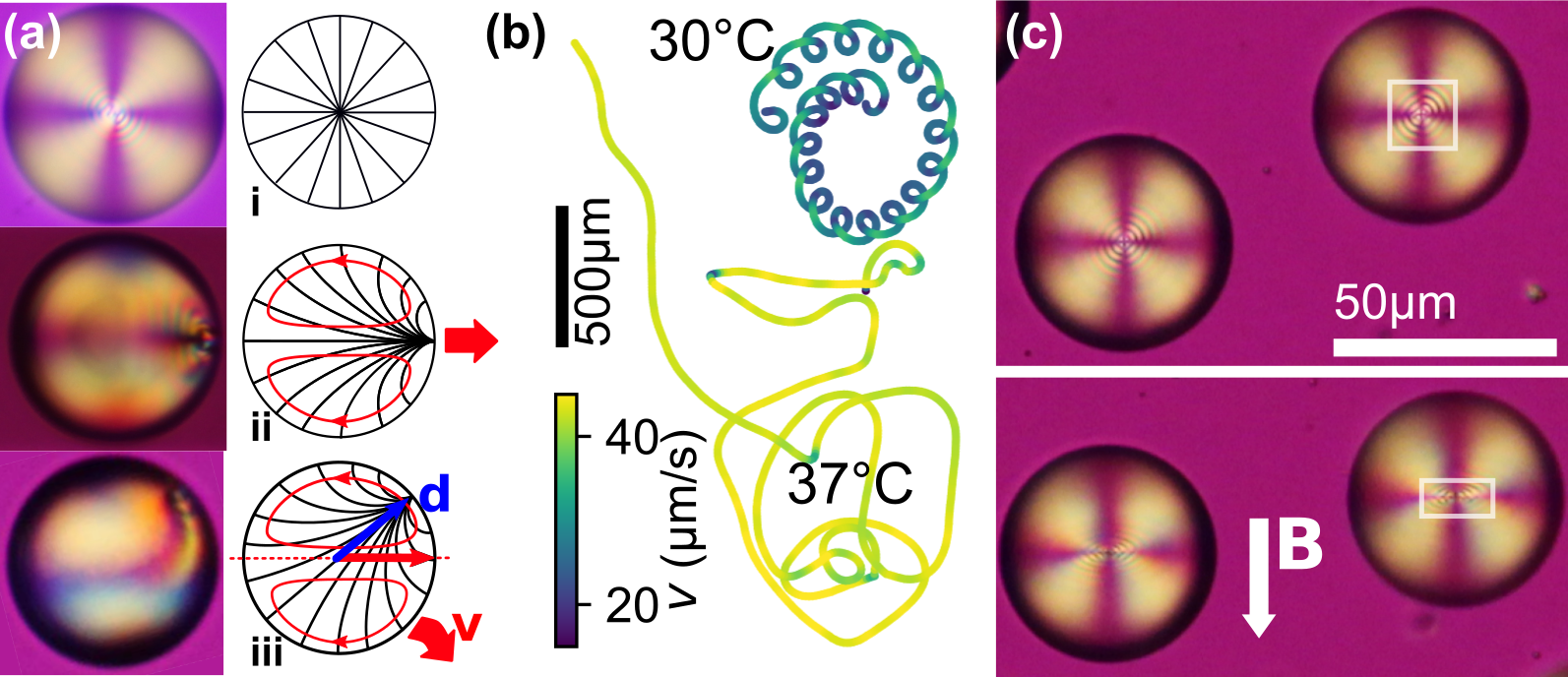}
    \caption{\textbf{Nematic self-propelling droplets.} \plb{a} Polarized images (left) and schematic of director fields and internal flow (right, black and red lines) for (i) droplet at rest, with a central defect, (ii) droplet swimming straight ahead, with anterior defect, (iii) droplet `curling' downward, with defect displaced upward~\cite{kruger2016_curling}. Arrows denote defect $\vec{d}$ and instantaneous velocity $\vec{v}$ relative to the droplet centroid. \plb{b} Quasi-ballistic non-nematic ($T = \SI{37}{\celsius}>T_\text{NI}$) vs. `curling' nematic 5CB swimmers ($T= \SI{30}{\celsius}<T_\text{NI}$). \plb{c} Deformation of the birefringence pattern in a resting droplet under a magnetic field of \SI{0.38}{\tesla}. The aspect ratio of the fringe pattern around the defect changes from $\approx1.0$ to $\approx2.0$ (white boxes). All droplets are of size $\approx\SI{50}{\um}$}
    \label{fig:fig1}
\end{figure}

The reference system of droplet swimmers we use~\cite{thutupalli2011_swarming,izri2014_selfpropulsion} comprises droplets of the thermotropic nematic oil 5CB in supramicellar solutions of the cationic surfactant TTAB~\cite{herminghaus2014_interfacial,maass2016_swimming}. Over time, these droplets solubilise into surfactant micelles~\cite{peddireddy2012_solubilization}, a process that locally increases the interfacial tension between oil and aqueous phases. Due to an advection-diffusion instability in the surfactant dynamics~\cite{michelin2013_spontaneous,izri2014_selfpropulsion,michelin2023_selfpropulsion}, higher modes in the interfacial tension arise with increasing P\'eclet number \Pen. \Pen\ increases with viscosity~\cite{hokmabad2021_emergence}, surfactant concentration~\cite{izzet2020_tunable} and droplet radius~\cite{suda2021_straightcurvilinear,ramesh2023_interfacial}. While small droplets at low surfactant concentration solubilise isotropically (interfacial mode $n=0$) and remain inactive, they self-propel as very persistent Brownian (or almost ballistic) swimmers  once \Pen\ exceeds the instability threshold for the dipolar $n=1$ mode~\cite{michelin2023_selfpropulsion}. We note that for even  higher \Pen, there is a complex interfacial mode spectrum and unsteady,  multi-modal dynamics~\cite{li2022_swimming,hokmabad2021_emergence}. In the present experiments, we keep to the quasi-ballistic regime by choosing a comparatively low surfactant concentration of \SI{7.5}{\wtpc} and droplets smaller than $\approx$\SI{60}{\um}. 

While the quasi-ballistic dynamics described above are typical for isotropic oils, like 5CB's isomer CB15~\cite{hokmabad2021_emergence} or diethylphtalate (DEP)~\cite{izzet2020_tunable}, nematic 5CB droplets swim in `curling' or meandering trajectories~\cite{kruger2016_curling,hokmabad2019_topological,suga2018_selfpropelled} due to nematoelastic effects~\cite{gennes1993_physics}, as follows: 
At the surfactant-covered oil-water interface, the liquid crystal anchors homeotropically~\cite{brake2003_effect}, requiring at least one defect of topological charge 1, typically a hedgehog, inside the droplet~\cite{kleman2006_topological,lopez-leon2011_drops}. At rest, the defect is centered (Fig.~\ref{fig:fig1}a-i). The fluid flow inside a moving droplet interacts with the director field~\cite{fernandez-nieves2007_topological,sengupta2013_liquid} and in this case advects the defect towards the anterior stagnation point (Fig.~\ref{fig:fig1}a-ii), and from there along the interface (Fig.~\ref{fig:fig1}a-iii). This asymmetry in the director field off the axis of motion causes an elastic restoring force and torque; since there are no external torques on the droplet, the torque has to be balanced by rotation. In principle, a constant torque would force the droplet onto a circle: however, as the droplets shed a chemorepulsive trail of spent fuel~\cite{jin2017_chemotaxis,moerman2017_solutemediated,hokmabad2022_chemotactic}, an additional chemorepulsive drift force forces them on elongated curling (Fig.~\ref{fig:fig1}b), meandering, or, in 3D, helical trajectories. When heated above 5CB's clearing temperature $T_c=\SI{34}{\celsius}$, the droplets swim quasi-ballistically (Fig.~\ref{fig:fig1}b). Notably, this spontaneous oscillatory instability does not require built-in chirality; however, chiral dopants have been employed to impose and control similar torques~\cite{lancia2019_reorientation}.

\subsection*{Nematic order deformations at rest in a magnetic field}

The dynamics described in the previous section apply to force and torque free conditions. We now proceed to the effects of an external magnetic field. 
5CB has a large diamagnetic anisotropy $\chi_a=\chi_\parallel-\chi_\perp$ stemming from its two phenyl groups~\cite{pauling1936_diamagnetic,gennes1993_physics}, such that the nematic director $\vec{n}$ has an affinity to align parallel to the field direction $\vec{H}$. De Gennes~\cite{gennes1993_physics} derives the following expressions for the free energy $F$ and the magnetic torque $\vec{\Gamma}_M$:
\begin{align}
F&=F_d-\frac12\chi_\perp{}H^2-\frac12\chi_a(\vec{n}\cdot\vec{H})^2\label{eq:F} \\
\vec{\Gamma}_M&=\chi_a(\vec{n}\cdot\vec{H})\vec{n}\times\vec{H}
\end{align}
Here, $F_d$ is the elastic energy stored in splay, bend and twist deformation of the director.
For the example of a droplet at rest, the free energy is minimized for a different director pattern under an applied magnetic field~\cite{stark1999_director}: we show this in Fig~\ref{fig:fig1}c by polarised images of inactive droplets under an in-plane magnetic field of \SI{0.38}{\tesla}. The aspect ratio of the birefringence pattern around the central defect changes from 1 to 2, indicating a significant reordering of the director in the magnetic field direction that competes against the radially symmetric splay imposed by the homeotropic boundary condition. 

Next, we study the dynamics of motile droplets in magnetic fields, depending on magnetic field strength, droplet size and degree of confinement.

\begin{figure}
    \centering
    \includegraphics[width=\columnwidth]{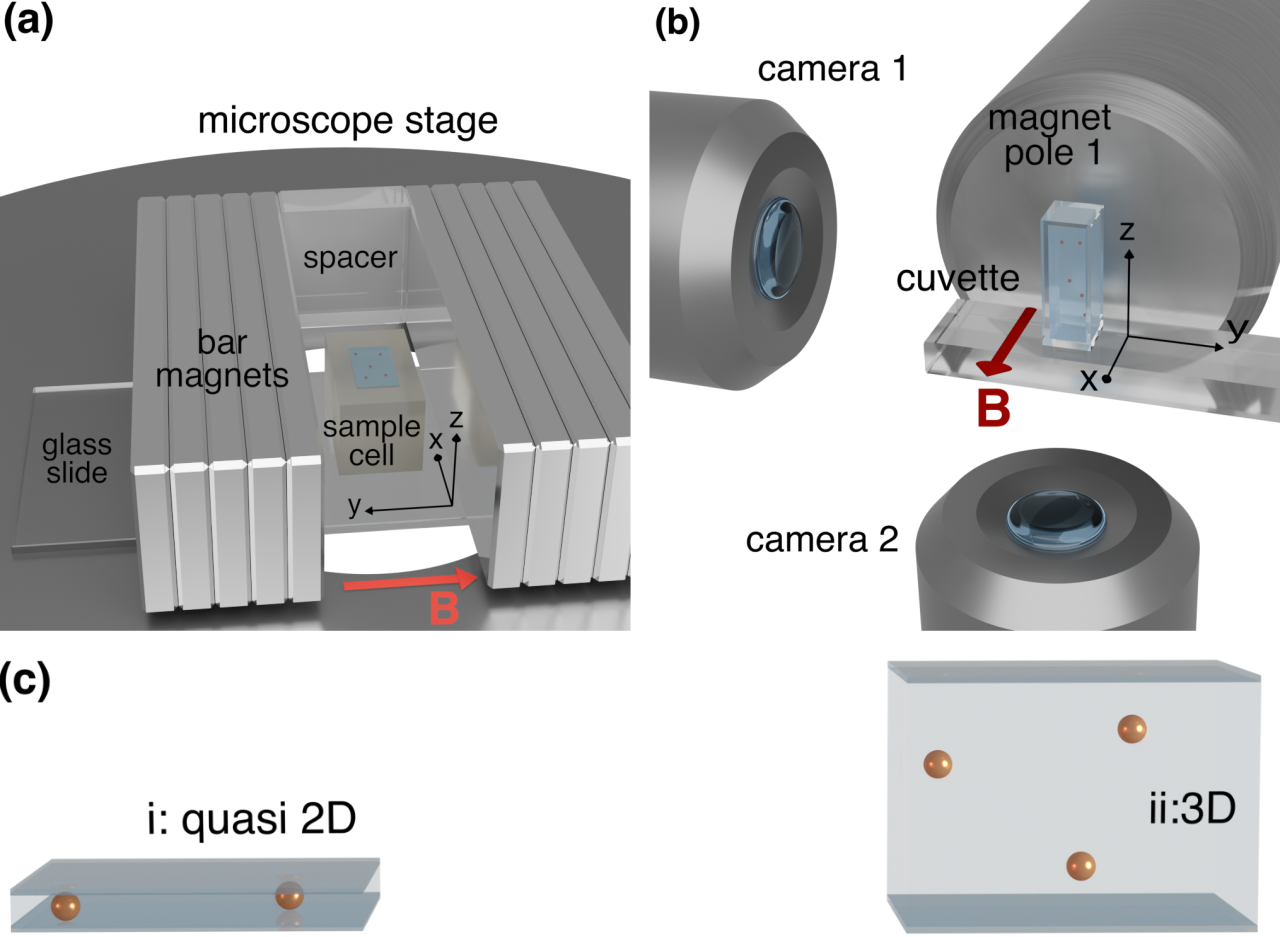}
    \caption{Setup concepts: \plb{a} for polarised high-resolution imaging with permanent bar magnets. \plb{b} for wide-field imaging between the yokes of an electromagnet. Magnet pole 2, on the other side of the sample holder in (b), and a second acrylic spacer in (a), are not shown for display purposes. \plb{c} confinement: Hele-Shaw for quasi-2D (i), bulk via density matched 3D (ii). Drawings not to scale.}
    \label{fig:setups}
\end{figure}

\begin{figure*}
    \centering    \includegraphics[width=\linewidth]{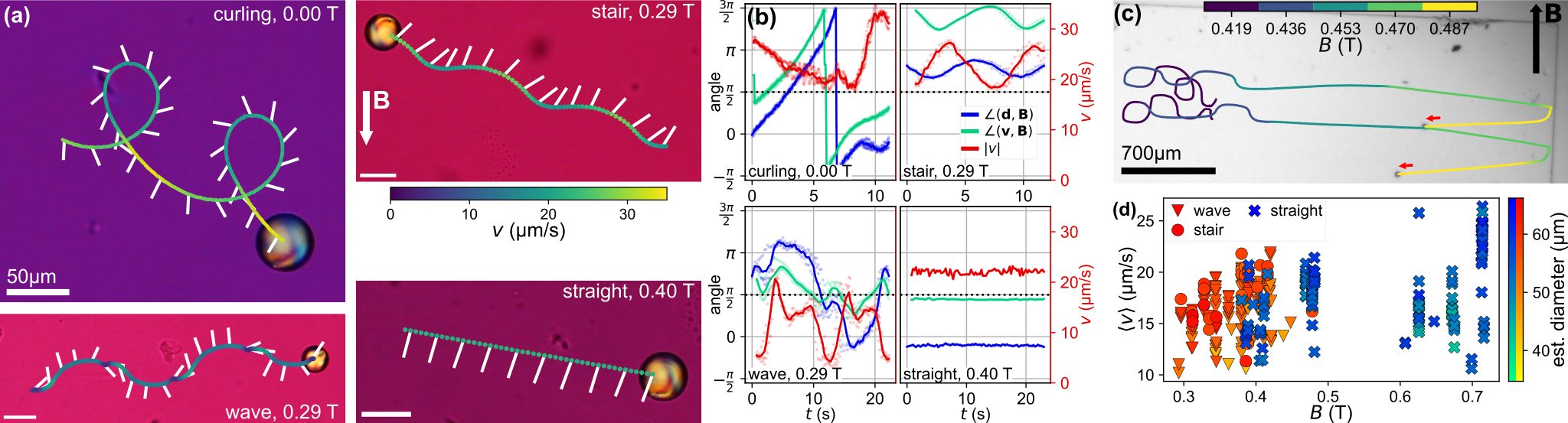}
    \caption{\textbf{Magnetotaxis in quasi 2D.}  \plb{a,b} Defect dynamics taken from polarized high resolution videmicrographs. Defect positions (white lines) were determined by frame-by-frame inspection. See also supporting videos S1-S4. \plb{c} The magnetic field is stepwise increased in 1 minute intervals. The dynamics change from oscillatory to straight around $B\approx \SI{0.43}{\tesla}$. \plb{d} Change from oscillatory to straight dynamics with increasing $B$, trajectories from 58 \SI{10}{\min}-experiments at constant $B$. 
    All experiments done at room temperature, \SI{22}{\celsius}. }
    \label{fig:fig3}
\end{figure*}

\begin{figure}
    \centering
\includegraphics[width=\columnwidth]{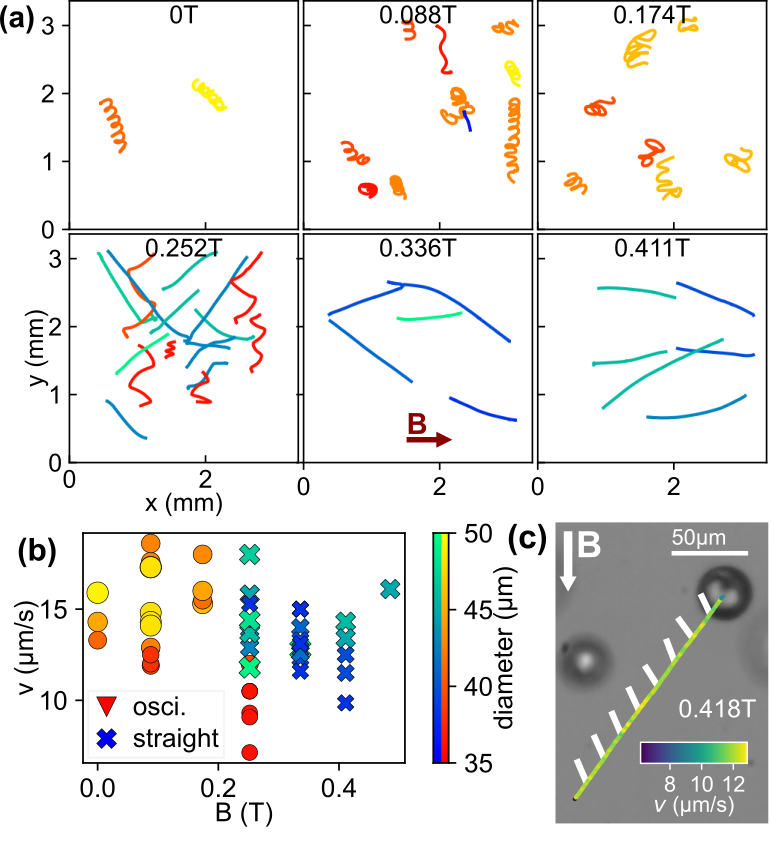}
    \caption{\plb{a} Magnetotaxis in force-free 3D, bottom view, with increasing $B$; colours distinguish between oscillatory and straight modes. With increasing field, we observe directional bias, elongation of the oscillation (see also Fig~\ref{fig:fig5}b,c) and a transition to straight motion.  (b) The transition to straight motion in the $v$/$B$ space. (c) Defect and speed orientation dynamics for a droplet magnetotaxing in 3D, from polarized microscopy, similar to Fig.~\ref{fig:fig3}a-straight. While the defect alignment is similar to the 2D case, the velocity orientation is different, i.e. not perpendicular to $\vec{B}$ any more.}
    \label{fig:fig4}
\end{figure}

\begin{figure*}
    \centering
\includegraphics[width=\linewidth]{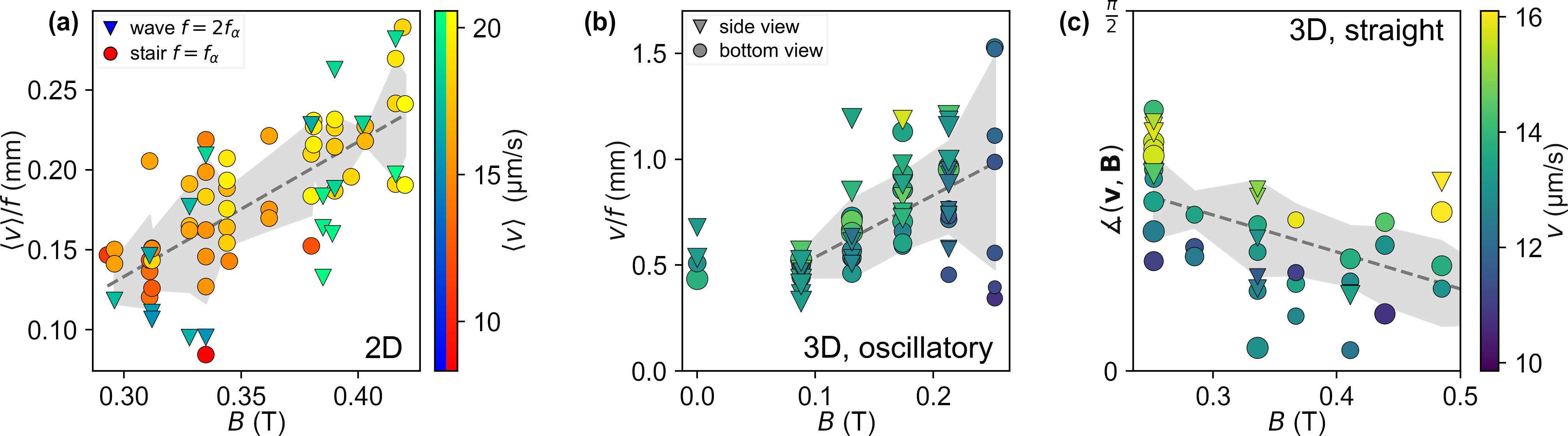}
    \caption{\plb{a} Oscillation length (average speed $v$/fundamental orientational frequency $f_\alpha$) versus $B$ for stair and wavy modes. The data for both modes collapse if $f_\alpha$ is doubled for the stair mode. \plb{b} The same for 3D motion, for all droplets oscillating in some manner. \plb{c} straight swimming droplets appear to align with $\vec{B}$ for increasing $B$, with more pronounced alignment for slower droplets. Dashed lines and error intervals are from linear regression fits and meant as guide to the eye.}
    \label{fig:fig5}
\end{figure*}

\subsection*{Setup}
Figure 1 shows an overview of the setup types and experimental cell geometries, with a definition of coordinate systems and field directions.

We used two types of setup to observe magnetotaxis: Type A for higher resolution polarized imaging (Fig.~\ref{fig:setups}a): here, we placed two symmetric stacks of rectangular neodymium (N40) bar magnets, held apart by acrylic spacers, on the stage of an Olympus IX-73 microscope equipped with polarization optics. The sample cell was centered in all three spatial directions between the magnets, with the field direction in the microscope viewing plane. We recorded videomicrographs with a colour Canon EOS 600D DSLR camera (video mode 1980$\times$1020px$^2$) or a monochrome CMOS camera (FLIR Grasshopper3, 2048$\times$2048px$^2$).

In type B (Fig.~\ref{fig:setups}b), we placed the sample between the poles of a GMW 3470 electromagnet, which allowed for seamless adjustment of the magnetic field  $B$, stereoscopic dual-camera  and wide-field observation, and a somewhat higher maximum field. Magnetic fields were measured with a Hall probe, the ambient temperature (RT) was \SI{22(1)}{\celsius}.  Wide-field microvideos (optionally stereoscopic) were recorded along the $y$ and/or $z$ axis using Grasshopper cameras and Navitar 12X zoom or commercial telephoto lenses.

To investigate the effects of confinement (Fig.~\ref{fig:setups}c), we varied the experimental cell geometry: either (i) a Hele-Shaw type cell of size $\SI{7}{\mm}\times\SI{4}{\mm}\times\SI{50}{\um}$, or, for bulk experiments, 
glass ($3\times3\times\SI{10}{\mm^3}$) cuvettes. In the latter case, we studied 
force-free off-boundary motion in the cuvette centre in a swimming medium matched to the density of 5CB ($\rho=\SI{1.022}{\g/\cm^3}$) by heavy water addition (ii). Droplets were produced in flow-focussing devices, either by in-house PDMS soft lithography~\cite{qin2010_soft} or in commercial glass chips (Micronit).

Droplet sizes, trajectories and velocities $\vec{v}$ as defined in Fig.~\ref{fig:fig1}a were determined by image processing algorithms using open source python and openCV libraries~\cite{bradski2000_opencv}. Sizes and positions were determined by binarization and contour analysis, followed by a nearest-neighbour analysis~\cite{crocker1996_methods} to reconstruct trajectories and instantaneous velocities.  We extracted the position and orientation of the defect $\vec{d}$ by frame-by-frame visual inspection of high resolution (20x-50x) polarized data.

\subsection{Quasi 2D confinement}
In 2D Hele-Shaw confinement, we observe the following behaviour with increasing magnetic field $B$ (Fig.~\ref{fig:fig3}): 
For $B=\SI{0}{\tesla}$, droplets follow \textit{curling} or meandering trajectories without directional bias, as published in~\cite{kruger2016_curling,suga2018_selfpropelled}.
Above $B\approx\SI{0.1}{\tesla}$, there is an orientational bias by the magnetic field, and we find two distinct oscillatory modes, which we call \textit{stair} and \textit{wave} in the following. At high fields, (Fig.~\ref{fig:fig3}c,d), all droplets swim \textit{straight}, i.e. non-oscillatory and ballistic, with a transitional region, where wavy, stair and straight modes coexist, around  $B\approx\SI{0.4}{\tesla}$.

We investigate the 4 modes in more detail in Fig.~\ref{fig:fig3}a,b (see also movies S1-S4). In panel (a), we plotted trajectories colour coded by speed on a polarized micrograph. The defect orientation is marked at regular intervals by white lines. In panel (b), we plotted the swimming direction $\measuredangle(\vec{v}, \vec{B})$ and the defect orientation $\measuredangle(\vec{d}, \vec{B})$, with respect to the magnetic field, as well as the speed $v$, averaged over 3 oscillations for curling, stair and wave, with $\vec{v},\vec{d}$ as defined in Fig.~\ref{fig:fig1}a. All four modes exhibit characteristic, regular behaviour in the dynamics of orientations and speeds plotted in panel (b). We note that, since the pseudo-vectorial liquid crystal director is insensitive to the polarity of $\vec{B}$, we did not distinguish between upward or downward field direction and assume $\vec{B}$ to point downwards without loss of generality. We found, within the experimental error, full top-down and left-right symmetry with respect to $\vec{B}$ for all our observations.

For the non-magnetotactic curling mode, there is a rather constant offset angle $\measuredangle(\vec{v}, \vec{d})$ between $\vec{v}$ and $\vec{d}$, while  $\vec{v}$ and $\vec{d}$ rotate freely in angular space. The speed $v$ decreases and increases somewhat around a trajectory self-intersection, probably due to chemorepulsive self-interactions~\cite{hokmabad2019_topological,hokmabad2022_chemotactic}. 

For the stair mode, $\vec{v}$ and $\vec{d}$ oscillate regularly, with the defect confined to one side of the droplet, i.e. never crossing the anterior pole. The trajectory shows a net orientation $\alpha = \pm(0.38\pm 0.04)\pi$ with the magnetic field, for all values of $B$ between 0.29 and \SI{0.475}{\tesla}. 

For the wave mode, $\vec{v}$ and $\vec{d}$ oscillate regularly as well, however with a much larger variation of the defect angle, which now crosses the anterior, such that the oscillations are symmetric in $\pm\vec{B}$. Due to this symmetry, the droplet moves on average perpendicular to $\vec{B}$ - we collected statistical evidence for this observation from widefield data of $\approx 70$ trajectories, which had a mean orientation of $(0.49\pm0.03)\pi$ with respect to $\pm\vec{B}$. Notably, when $\measuredangle(\vec{d}, \vec{B})$ and $\measuredangle(\vec{v}, \vec{B})$ intersect at an angle $\pi/2$, the droplet speed is minimal, here $\SI{2}{\um/s}$ compared to a maximum speed of \SI{20}{\um/s}, which suggests a maximum of viscous dissipation at this point.

While both stair and wave mode coexist in a range of magnetic fields between 0.29 and \SI{0.475}{\tesla}, we found in 10-minute measurements during which the droplets shrank by about 10\% that larger droplets starting in stair mode switched to the wave mode, below an estimated size threshold of around \SI{50}{\um}, with initial and final droplet sizes estimated from separate high resolution imaging. While this observation suggests a dependence on either diameter $d$ or degree of confinement $H/d$, we note that the size determination from wide-field data is not accurate enough to confirm this with sufficient statistics at present.

We make the following observations regarding the straight mode, as shown for an example at $B=\SI{0.4}{\tesla}$ in Fig.~\ref{fig:fig3}. All orientations and the speed are constant. The defect is oriented considerably off the polar axis of the droplet, $\measuredangle (\vec{v},\vec{d})\approx 0.54\pi$. This suggests a significant nematoelastic torque, which must be entirely compensated by the diamagnetic torque exerted by the magnetic field to sustain ballistic motion.

We also observe that the defect orients close to the direction of $\vec{B}$, $\measuredangle (\vec{B},\vec{d})\approx 0.1\pi$, which is quite similar to its maximum deviation from the polar axis in the wave mode (Fig.~\ref{fig:fig3}c, bottom row), and the droplet speed is comparable to the maximum speed in the wave mode, $\approx\SI{20}{\um/\s}$. We hypothesize that these orientations correspond to a state of minimal dissipation, and that the oscillations in the wave mode are a consequence of the torque imposed by the lower magnetic field not being sufficient to compensate the nematoelastic one. 

\subsection{Density matched 3D}

We have found in the $z$ confined experiments analyzed above that the confinement significantly influences the modes of motion, both by breaking the spatial symmetry, by constraining the defect dynamics to the $xy$ plane, and, presumably, also by the effect of confinement on the flow fields generated by the droplet's propulsion. It is therefore of interest to compare these experiments to a case without confinement, as provided by droplets in a mesoscopic quarz cuvette in a density matched swimming medium, minimizing the influence of gravity and container boundaries. 

Fig.~\ref{fig:fig4}a,b summarizes a series of experiments with $B$ between 0 and $\SI{0.411}{\tesla}$, with panel (a) showing trajectories of droplets within an estimated size range between 35 and $\SI{50}{\um}$. We note that we here used relatively polydisperse samples and that due to the nature of wide field videomicrography in 3D these size estimates have a large uncertainty of $\approx 20$\%. 

Under these caveats, we note the following trends: At $B=0$, the droplet move in helices without directional bias, as previously published~\cite{kruger2016_curling}. With increasing $B$, here shown for $B\geq\SI{0.088}{\tesla}$, this helical motion is biased perpendicularly to $\vec{B}$. At intermediate fields, similar to the 2D case, there is a coexistence of oscillatory and straight motion (Fig.~\ref{fig:fig4}b). However, the transition appears to occur at a somewhat lower field compared to 2D:  around $B=\SI{0.25}{\tesla}$  droplets of smaller appearance ($<\SI{40}{\um}$) still oscillate while bigger ones swim straight, at an angle 
$\measuredangle (\vec{v},\vec{B})$ around $\pi/4$. This angle appears to reduce with increasing $B$, such that the droplets tend to align with $\vec{B}$ at $B=\SI{0.411}{\tesla}$, in contrast to the quasi 2D case, where the straight mode favours perpendicular alignment. To statistically investigate this trend we collected droplet orientations from several experimental runs with a total of 44 recorded straight trajectories, and plotted $\measuredangle(\vec{v},\vec{B})$, versus $B$ in Fig.~\ref{fig:fig5}c. Here, $\measuredangle(\vec{v},\vec{B})$ indeed decreases with increasing $B$, with slower droplets appearing to favour stronger alignment.
We pursued this further by recording a droplet in a density matched 3D ambient medium at 20x magnification under polarized microscopy (Fig.~\ref{fig:fig4}c) and estimated the defect orientation  (white lines) by frame-by-frame visual inspection. Notably, this defect orientation with respect to $\vec{B}$ is close to the 2D case, $\measuredangle (\vec{B},\vec{d})\approx 0.15\pi$. However, the angle between defect and droplet orientation is much smaller in this projection, $\measuredangle (\vec{v},\vec{d})\approx 0.34\pi$, corresponding to the closer alignment of the droplet motion with $\vec{B}$.

Finally, we investigate the transition to straight motion with increasing $B$ in its effect on the oscillation for both quasi 2D and 3D cases. From the trajectories plotted in Fig.~\ref{fig:fig3}c and Fig.~\ref{fig:fig4}a, we find that the oscillations elongate or coarsen in space when the magnetic field approaches the transition to straight motion.
To probe this lengthscale, we extracted the dominant frequency $f_\alpha$ of the trajectory orientation via fast Fourier transformation (FFT) for 2D wave and stair (Fig.~\ref{fig:fig5}a), as well as 3D helical modes (Fig.~\ref{fig:fig5}b) and divided them by the estimated droplet speed as a measure of characteristic length. For both geometries, this length increases roughly linearly with $B$, and it is about twice as large in 3D as in quasi 2D. The data for wave and stair modes collapse if $f_\alpha$ is doubled for the wave mode - a consequence of the fact that the dominant frequency extracted by FFT encompasses two symmetric half cycles to both sides of the polar axis for the wave case, and only one for stair (note also that, correspondingly, the absolute speed in Fig.~\ref{fig:fig3}b oscillates at twice the frequency of defect and speed for the wave mode).

\section*{Discussion}
Based on the results above, we propose the following working hypothesis: a steady state of motion corresponds to minimizing a free energy functional depending on diamagnetic and nematoelastic energy and viscous dissipation due to the inner and outer flow fields~\cite{tang2020_minimization}.
Apparently this energy is minimized for an off-axis defect displacement, as seen for the curling state at zero field, and is therefore not a state of neutral torque. If the external torque by the magnetic field is strong enough to compensate the nematoelastic one, straight motion is possible. If not, the residual torque will rotate the droplet away, causing defect oscillation and rotation. The deformation of the director field, as well as the internal flow field, depend sensitively on the droplet size, as well as its speed, such that we can expect different behaviours under variation of these parameters. However, since all these quantities are coupled, further insight will require a numerical analysis of the problem and an at least qualitative comparison to the defect patterns gathered from polarized microscopy~\cite{bahr2021_lattice, hokmabad2019_topological}. 

The switch between stair and wave might be motivated as follows: In strong confinement, the droplet is not allowed to rotate up or down, as it would bump into the cell boundaries. In consequence, the defect has to rotate in the cell midplane as well. As noted above, there appears to be a strong dissipative energy barrier for the defect to cross the droplet anterior, such that it is trapped on one side of the droplet by both the $z$ confinement and this barrier. If the confinement is relaxed, e.g. when the droplet shrinks, the defect can escape to the other side of the droplet, leading to the symmetric wave oscillation. These confinement effects would naturally not apply to the 3D case, where our stereoscopic experiments indicate a rectified helical motion for moderate magnetic fields.

Interestingly, and related in spirit, the diamagnetic alignment of a bulk liquid crystal has been used to rectify the flows in a dynamically coupled active nematic layer~\cite{guillamat2016_control}.

Our results offer a promising demonstration of how the interplay of topology and symmetry broken by activity~\cite{bowick2022_symmetry} can lead to a rich spectrum of controllable motile states in microswimmers that can be harnessed for technological application.

\section*{Acknowledgments}
We thank Thomas Zijlstra and Gert-Wim Bruggert for invaluable technical advice and support and Christian Bahr for continuously sharing his advice and expertise on liquid crystals and active interfaces.

\bibliographystyle{unsrt}

\end{document}